\documentclass[preprint2,times]{aastex631}

\PassOptionsToPackage{dvipsnames,svgnames,x11names}{xcolor}
\usepackage{tabularx}

\usepackage{amsmath}  
\usepackage{amssymb}  
\usepackage[T1]{fontenc}
\usepackage{xspace}
\usepackage{multirow}
\usepackage{xcolor}

\defcitealias{kenworthy_salt3_2021}{K21}

\usepackage{newunicodechar,graphicx}
\DeclareRobustCommand{\okina}{%
  \raisebox{\dimexpr\fontcharht\font`A-\height}{%
    \scalebox{0.8}{`}%
  }%
}
\newunicodechar{ʻ}{\okina}

\newif{\ifchangetext}
\changetextfalse

\InputIfFileExists{\jobname-options}

\ifchangetext
  
  \newcommand{\changenote}[1]{\textcolor{blue}{ \bf #1}}x
\else
  
  \newcommand{\changenote}[1]{}
\fi

\hypersetup{
    colorlinks=True,
    citecolor=black,
    linkcolor=black,
    filecolor=magenta,      
    urlcolor=cyan,
}

\def\arcsec{\ensuremath{^{\prime\prime}}}

\newcommand{\lensedsn}{SN\,Encore\xspace}

\newcommand{\hst}{\textit{HST}\xspace}

\newcommand{\webb}{\textit{JWST}\xspace}

\newcommand{\STScI}{Space Telescope Science Institute, Baltimore, MD 21218, USA}
\newcommand{\JHU}{Physics and Astronomy Department, Johns Hopkins University, Baltimore, MD 21218, USA}
\newcommand{\UOArizona}{Department of Astronomy/Steward Observatory, University of Arizona, 933 N. Cherry Avenue, Tucson, AZ 85721, USA}
\newcommand{\ISEF}{ISEF International Fellowship}
\newcommand{\NEF}{NASA Einstein Fellow}

\usepackage{xspace}

\begin{document}

\title{Lensed Type Ia Supernova ``Encore'' at $\mathbf{z=2}$: The First Instance of Two Multiply-Imaged Supernovae in the Same Host Galaxy}

\author[0000-0002-2361-7201
]{J.~D.~R.~Pierel}
\correspondingauthor{J.~D.~R.~Pierel} 
\email{jpierel@stsci.edu}
\altaffiliation{\NEF}
\affiliation{\STScI}

\author[0000-0001-7769-8660]{A.~B.~Newman} 
\affiliation{Observatories of the Carnegie Institution for Science, 813 Santa Barbara St., Pasadena, CA 91101, USA}

\author[0000-0002-2376-6979]{S.~Dhawan} 
\affiliation{Institute of Astronomy and Kavli Institute for Cosmology, University of Cambridge, Madingley Road, Cambridge CB3 0HA, UK
}

\author[0000-0002-4267-9344]{M.~Gu} 
\affiliation{Department of Physics, The University of Hong Kong, Pok Fu Lam, Hong Kong
}

\author[0000-0002-7593-8584]{B.~A.~Joshi} 
\affiliation{\JHU}

\author[0009-0005-5008-0381]{T.~Li} 
\affiliation{Institute of Cosmology and Gravitation, University of Portsmouth, Burnaby Road, Portsmouth, PO1 3FX, UK
}

\author[0000-0003-2497-6334]{S.~Schuldt} 
\affiliation{Dipartimento di Fisica, Universit\`a  degli Studi di Milano, via Celoria 16, I-20133 Milano, Italy}
\affiliation{INAF - IASF Milano, via A. Corti 12, I-20133 Milano, Italy
}

\author[0000-0002-7756-4440]{L.~G.~Strolger} 
\affiliation{\STScI}

\author[0000-0001-5568-6052]{S.~H.~Suyu
} 
\affiliation{Technical University of Munich, TUM School of Natural Sciences, Physics Department, James-Franck-Str.~1, 85748 Garching, Germany}
\affiliation{Max-Planck-Institut f\"ur Astrophysik, Karl-Schwarzschild-Str. 1, D-85748 Garching, Germany 
}
\affiliation{Institute of Astronomy and Astrophysics, Academia Sinica, 11F of ASMAB, No.~1, Section 4, Roosevelt Road, Taipei 10617,Taiwan}
\author[0000-0001-6052-3274]{G.~B.~Caminha} 
\affiliation{Technical University of Munich, TUM School of Natural Sciences, Physics Department, James-Franck-Str.~1, 85748 Garching, Germany}
\affiliation{Max-Planck-Institut f\"ur Astrophysik, Karl-Schwarzschild-Str. 1, D-85748 Garching, Germany 
}

\author[0000-0003-3329-1337]{S.~H.~Cohen}
\affiliation{School of Earth and Space Exploration, Arizona State University,
Tempe, AZ 85287-1404, USA}

\author[0000-0001-9065-3926]{J.~M.~Diego}
\affiliation{Instituto de Fisica de Cantabria (CSIC-C). Avda. Los Castros s/n. 39005 Santander, Spain}

\author[0000-0002-9816-1931]{J.~C.~J.~D\'Silva}
\affiliation{International Centre for Radio Astronomy Research (ICRAR) and the
International Space Centre (ISC), The University of Western Australia, M468,
35 Stirling Highway, Crawley, WA 6009, Australia}
\affiliation{ARC Centre of Excellence for All Sky Astrophysics in 3 Dimensions
(ASTRO 3D), Australia}

\author[0000-0002-5085-2143]{S.~Ertl} 
\affiliation{Technical University of Munich, TUM School of Natural Sciences, Physics Department, James-Franck-Str.~1, 85748 Garching, Germany}
\affiliation{Max-Planck-Institut f\"ur Astrophysik, Karl-Schwarzschild-Str. 1, D-85748 Garching, Germany 
}

\author[0000-0003-1625-8009
]{B.~L.~Frye}
\affiliation{\UOArizona}

\author[0000-0002-9512-3788]{G.~Granata} 
\affiliation{Dipartimento di Fisica, Universit\`a  degli Studi di Milano, via Celoria 16, I-20133 Milano, Italy}
\affiliation{Dipartimento di Fisica e Scienze della Terra, Universit\`a degli Studi di Ferrara, via Saragat 1, I-44122 Ferrara, Italy}
\affiliation{INAF - IASF Milano, via A. Corti 12, I-20133 Milano, Italy
}
\author[0000-0002-5926-7143]{C.~Grillo} 
\affiliation{Dipartimento di Fisica, Universit\`a  degli Studi di Milano, via Celoria 16, I-20133 Milano, Italy}
\affiliation{INAF - IASF Milano, via A. Corti 12, I-20133 Milano, Italy
}
\author[0000-0002-6610-2048]{A.~M.~Koekemoer}
\affiliation{\STScI}

\author[0000-0002-8184-5229]{J.~Li}
\affiliation{International Centre for Radio Astronomy Research (ICRAR) and the
International Space Centre (ISC), The University of Western Australia, M468,
35 Stirling Highway, Crawley, WA 6009, Australia}

\author[0000-0003-0429-3579]{A.~Robotham}
\affiliation{International Centre for Radio Astronomy Research (ICRAR) and the
International Space Centre (ISC), The University of Western Australia, M468,
35 Stirling Highway, Crawley, WA 6009, Australia}

\author[0000-0002-7265-7920]{J.~Summers} 
\affiliation{School of Earth and Space Exploration, Arizona State University,
Tempe, AZ 85287-1404, USA}

\author[0000-0002-8460-0390]{T.~Treu} 
\affiliation{Physics and Astronomy Department, University of California, Los Angeles CA 90095
}

\author[0000-0001-8156-6281]{R.~A.~Windhorst} 
\affiliation{School of Earth and Space Exploration, Arizona State University,
Tempe, AZ 85287-1404, USA}

\author[0000-0002-0350-4488]{A.~Zitrin} 
\affiliation{Department of Physics, Ben-Gurion University of the Negev, P.O. Box 653, Beer-Sheva, 84105, Israel}


\author[0000-0002-2350-4610]{S.~Agarwal} 
\affiliation{Department of Astronomy, University of California, 501 Campbell Hall \#3411, Berkeley, CA 94720, USA
}

\author[0009-0008-1965-9012
]{A.~Agrawal} 
\affiliation{Department of Astronomy, University of Illinois, Urbana, IL 61801, USA
}

\author[0000-0001-5409-6480]{N.~Arendse} 
\affiliation{Oskar Klein Centre, Department of Physics, Stockholm University, SE-106 91 Stockholm, Sweden
}

\author[0000-0001-8156-6281]{S.~Belli} 
\affiliation{Dipartimento di Fisica e Astronomia, Universit\'a di Bologna, via Piero Gobetti 93/2, 40129 Bologna, Italy}

\author[0000-0003-4625-6629]{C.~Burns} 
\affiliation{Observatories of the Carnegie Institution for Science, 813 Santa Barbara St., Pasadena, CA 91101, USA}

\author[0000-0002-2468-5169]{R.~Ca\~nameras} 
\affiliation{Max-Planck-Institut f\"ur Astrophysik, Karl-Schwarzschild-Str. 1, D-85748 Garching, Germany 
}
\affiliation{Technical University of Munich, TUM School of Natural Sciences, Physics Department, James-Franck-Str.~1, 85748 Garching, Germany}

\author[0000-0001-6711-8140]{S.~Chakrabarti} 
\affiliation{Department of Physics and Astronomy, University of Alabama, 301 Sparkman Drive Huntsville, AL 35899
}

\author[0000-0003-1060-0723]{W.~Chen} 
\affiliation{Department of Physics,
Oklahoma State University, 145 Physical Sciences Bldg, Stillwater, OK
74078, USA}

\author[0000-0001-5564-3140]{T.~E.~Collett} 
\affiliation{Institute of Cosmology and Gravitation, University of Portsmouth, Burnaby Road, Portsmouth, PO1 3FX, UK
}

\author[0000-0003-4263-2228]{D.~A.~Coulter} 
\affiliation{\STScI} 

\author[0000-0001-7782-7071]{R.~S.~Ellis} 
\affiliation{University College London, Gower St, London WC1E 6BT}

\author[0000-0003-0209-674X]{M.~Engesser}
\affiliation{\STScI}

\author[0000-0002-7460-8460]{N.~Foo} 
\affiliation{School of Earth and Space Exploration, Arizona State University,
Tempe, AZ 85287-1404, USA}

\author[0000-0003-2238-1572]{O.~D.~Fox} 
\affiliation{\STScI}

\author[0000-0002-8526-3963]{C.~Gall} 
\affiliation{DARK, Niels Bohr Institute, University of Copenhagen, Jagtvej 155, 2200 Copenhagen, Denmark
}

\author[0000-0003-3418-2482]{N.~Garuda} 
\affiliation{Department of Astronomy/Steward Observatory, University of Arizona, 933 N. Cherry Avenue, Tucson, AZ 85721, USA
}

\author[0000-0003-3703-5154]{S.~Gezari}
\affiliation{\STScI}

\author[0000-0001-6395-6702]{S.~Gomez}
\affiliation{\STScI}

\author[0000-0002-3254-9044]{K.~Glazebrook} 
\affiliation{Centre for Astrophysics and Supercomputing, Swinburne University of Technology, PO Box 218, Hawthorn, VIC 3122, Australia
}

\author[0000-0002-4571-2306]{J.~Hjorth} 
\affiliation{DARK, Niels Bohr Institute, University of Copenhagen, Jagtvej 155, 2200 Copenhagen, Denmark
}

\author[0000-0001-8156-0330]{X.~Huang} 
\affiliation{Department of Physics \& Astronomy, University of San Francisco, San Francisco, CA 94117-1080
}
\affiliation{Kavli Institute for Cosmological Physics, University of Chicago, Chicago, IL 60637, USA
}

\author[0000-0001-8738-6011]{S.~W.~Jha} 
\affiliation{Department of Physics and Astronomy, Rutgers University, 136 Frelinghuysen Road, Piscataway, NJ 08854, USA
}

\author[0000-0001-9394-6732]{P.~S.~Kamieneski} 
\affiliation{School of Earth and Space Exploration, Arizona State University,
Tempe, AZ 85287-1404, USA}

\author[0000-0003-3142-997X]{P.~Kelly} 
\affiliation{Minnesota Institute for Astrophysics, 116 Church St SE, Minneapolis, MN 55455
}

\author[0000-0003-2037-4619]{C.~Larison} 
\affiliation{Department of Physics and Astronomy, Rutgers University, 136 Frelinghuysen Road, Piscataway, NJ 08854, USA
}

\author[0000-0003-3030-2360]{L.~A.~Moustakas} 
\affiliation{Jet Propulsion Laboratory, California Institute of Technology, 4800 Oak Grove Dr, Pasadena, CA 91109
}

\author[0000-0002-2282-8795]{M.~Pascale} 
\affiliation{Department of Astronomy, University of California, 501 Campbell Hall \#3411, Berkeley, CA 94720, USA
}

\author[0000-0002-2807-6459]{I.~P\'erez-Fournon} 
\affiliation{Instituto de Astrof\'{\i}sica de Canarias, V\'{\i}a L\'actea, 38205 La Laguna, Tenerife, Spain.} 
\affiliation{Universidad de La Laguna, Departamento de Astrof\'{\i}sica,  38206 La Laguna, Tenerife, Spain.
}

\author[0000-0003-4743-1679]{T.~Petrushevska}
\affiliation{Center for Astrophysics and Cosmology, University of Nova Gorica, Vipavska 11c, 5270 Ajdov\v{s}\v{c}ina, Slovenia.
}

\author[0000-0002-5391-5568]{F.~Poidevin} 
\affiliation{Instituto de Astrof\'{\i}sica de Canarias, V\'{\i}a L\'actea, 38205 La Laguna, Tenerife, Spain.}
\affiliation{Universidad de La Laguna, Departamento de Astrof\'{\i}sica,  38206 La Laguna, Tenerife, Spain.
}

\author[0000-0002-4410-5387]{A.~Rest} 
\affiliation{\STScI}
\affiliation{\JHU}

\author[0000-0002-9301-5302]{M.~Shahbandeh} 
\affiliation{\STScI}

\author[0000-0002-5558-888X]{A.~J.~Shajib} 
\altaffiliation{\NEF}
\affiliation{Department of Astronomy \& Astrophysics, The University of Chicago, Chicago, IL 60637, USA
}
\affiliation{Kavli Institute for Cosmological Physics, University of Chicago, Chicago, IL 60637, USA
}

\author[0000-0003-2445-3891]{M.~Siebert} 
\affiliation{\STScI}

\author[0000-0002-0385-0014]{C.~Storfer} 
\affiliation{Institute for Astronomy, University of Hawaii, Honolulu, HI 96822-1897}

\author{M.~Talbot} 
\affiliation{Jet Propulsion Laboratory, California Institute of Technology, 4800 Oak Grove Dr, Pasadena, CA 91109
}

\author[0000-0001-5233-6989]{Q.~Wang} 
\affiliation{\JHU}

\author[0000-0002-4043-9400]{T.~Wevers}
\affiliation{\STScI}

\author[0000-0002-0632-8897]{Y.~Zenati}
\altaffiliation{\ISEF}
\affiliation{\JHU}
\affiliation{\STScI}

\begin{abstract}
A bright ($m_{\rm F150W,AB}$\,=\,24 mag), $z=1.95$ supernova (SN) candidate was discovered in {\it JWST}/NIRCam imaging acquired on 2023 November 17.  The SN is quintuply-imaged as a result of strong gravitational lensing by a foreground galaxy cluster, detected in three locations, and remarkably is the second lensed SN found in the same host galaxy. The previous lensed SN was called ``Requiem'', and therefore the new SN is named ``Encore''. This makes the MACS J0138.0$-$2155 cluster the first known system to produce more than one multiply-imaged SN. Moreover, both SN Requiem and \lensedsn are Type Ia SNe (SNe\,Ia), making this the most distant case of a galaxy hosting two SNe\,Ia. Using parametric host fitting, we determine the probability of detecting two SNe\,Ia in this host galaxy over a $\sim10$ year window to be $\approx3\%$. These observations have the potential to yield a Hubble Constant ($H_0$) measurement with $\sim10\%$ precision, only the third lensed SN capable of such a result, using the three visible images of the SN. Both SN Requiem and SN Encore have a fourth image that is expected to appear within a few years of $\sim2030$, providing an unprecedented baseline for time-delay cosmography. 

\end{abstract}

\section{Introduction}
\label{sec:intro}
Strong gravitational lensing can cause multiple images of a background source to appear, as light propagating along different paths are focused by the gravity of a foreground galaxy or galaxy cluster \citep[called the ``lens''; e.g.,][]{narayan_lectures_1997}. Such a phenomenon requires chance alignment between the observer, the background source, and the lens. If the multiply-imaged source has variable brightness then depending on the relative geometrical and gravitational potential differences of each path, the source images will typically appear delayed by hours to weeks (for galaxy-scale lenses, $M\lesssim10^{12}M_\odot$) or months to years \citep[for cluster-scale lenses, $M\gtrsim10^{13}M_\odot$; e.g.,][]{oguri_strong_2019}. 

\begin{figure*}[!ht]
    \centering
        \includegraphics[trim={0cm 4cm 0cm 0cm},clip,width=.99\linewidth]{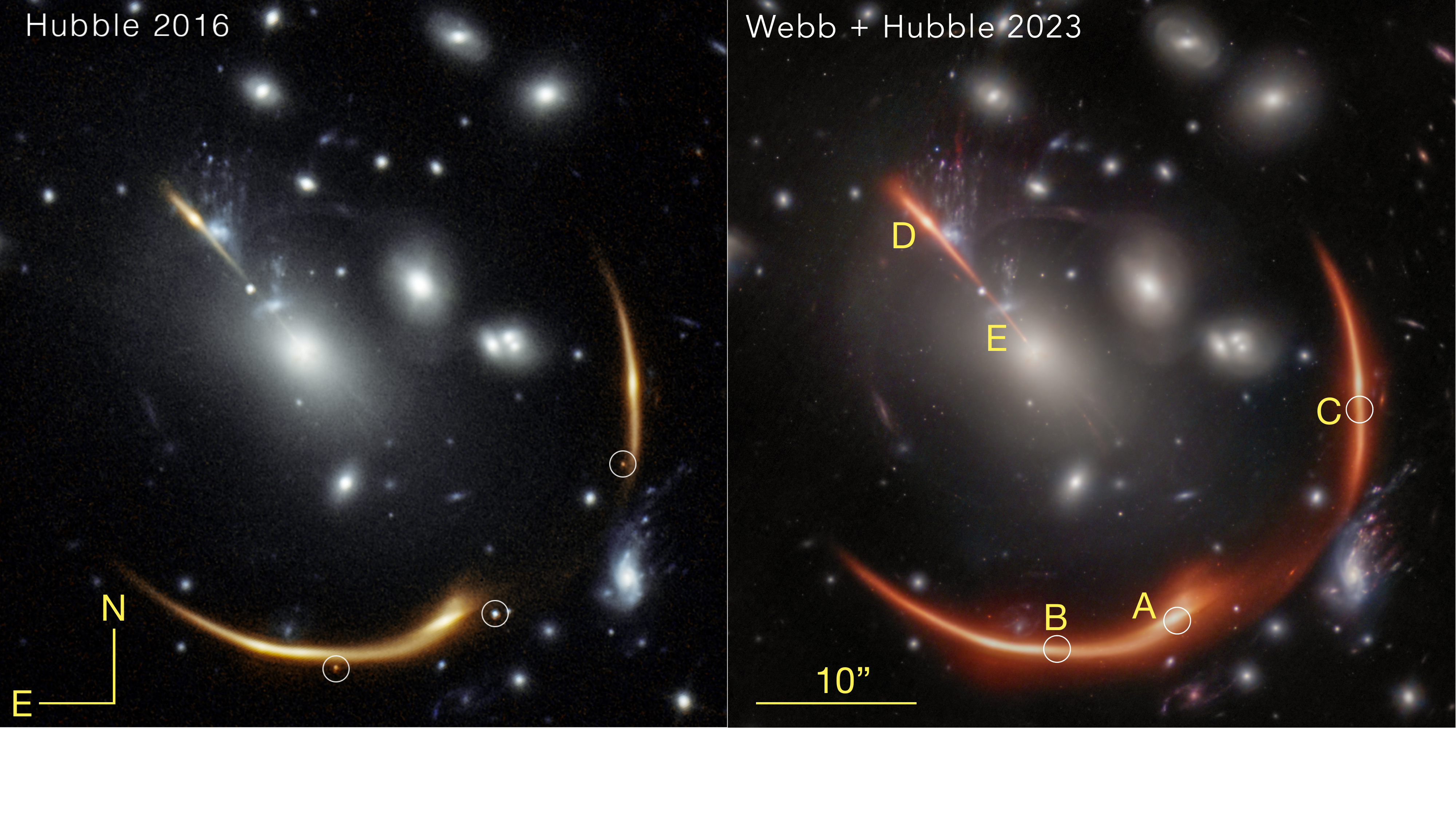}
    \caption{ {\it Left}: {\it HST} WFC3/IR two-color image in the region of MACS J0138.0$-$2155 from 2016, using F105W (blue) and F160W (orange). SN Requiem is marked in its three visible image positions by white circles, notably absent in $2023$. {\it Right:} Combined \textit{JWST}/NIRCam and \textit{HST}/WFC3 color image from programs 6549 and 16264 (Table \ref{tab:obs}). The filters used are F105W+F115W+F125W (blue), F150W+F160W+F200W (green), and F277W+F356W+F444W (red). The images were drizzled to $0\farcs02/\rm{pix}$, and the image scale and orientation are as shown. The three detected image positions of \lensedsn are circled, but it is not visible at this scale, and all five images of the host galaxy are labeled (A-E). For a zoom in of each of the detected image positions, see Figure \ref{fig:cutouts}  (Image Credit: STScI, A. Koekemoer, T. Li).}
    \label{fig:color_im}
\end{figure*}

Precise measurements of this ``time delay'' yield a direct distance measurement to the lens system that constrains the Hubble constant ($H_0$) in a single step \citep[e.g.,][]{refsdal_possibility_1964,linder_lensing_2011,paraficz_gravitational_2009,treu_time_2016,grillo_measuring_2018,grillo_accuracy_2020,birrer_time-delay_2022,treu_strong_2022,kelly_constraints_2023,grillo_cosmography_2024,suyu_strong_2024}. While this has been accomplished with quasars \citep[e.g.,][]{kundic_robust_1997,schechter_quadruple_1997,burud_time_2002,hjorth_time_2002,vuissoz_cosmograil_2008,suyu_dissecting_2010,tewes_cosmograil_2013,bonvin_h0licow_2017,bonvin_cosmograil_2018,bonvin_cosmograil_2019,birrer_h0licow_2019,chen_sharp_2019,wong_h0licow_2020,birrer_tdcosmo_2020,shajib_strides_2020,shajib_tdcosmo_2023}, there is much discussion about the advantages of using supernovae (SNe) instead to leverage a variety of valuable characteristics such as predictable evolution, simplicity of observations, standardizable brightness (for Type Ia SNe [SNe\,Ia]), and mitigated microlensing effects \citep[e.g.,][]{foxley-marrable_impact_2018,goldstein_precise_2018,pierel_turning_2019,huber_holismokes_2021,ding_improved_2021,pierel_projected_2021,birrer_hubble_2022,chen_jwst_2024,pascale_sn_2024,pierel_jwst_2024}. The sample of strongly lensed SNe has grown over the last decade to include two galaxy-scale systems \citep{goobar_iptf16geu_2017,goobar_uncovering_2023,pierel_lenswatch_2023} and five cluster-scale systems \citep{kelly_multiple_2015,rodney_gravitationally_2021,chen_jwst-ers_2022,kelly_strongly_2022,frye_jwst_2024}, though only SN Refsdal \citep{kelly_constraints_2023,kelly_magnificent_2023} and SN H0pe \citep{chen_jwst_2024,frye_jwst_2024,pascale_sn_2024,pierel_jwst_2024}  have had sufficiently long time-delays and well-sampled light curves for $H_0$ constraints with relatively small uncertainty. SN Refsdal yielded a $\sim$6\% measurement of $H_0$ in flat $\Lambda$CDM cosmology \citep[64.8$_{-4.3}^{+4.4}$ or 66.6$_{-3.3}^{+4.1 }$ km s$^{-1}$ Mpc$^{-1}$, depending on lens model weights;][]{kelly_constraints_2023}, 
and also $65.1_{-3.4}^{+3.5 }$ km s$^{-1}$ Mpc$^{-1}$ in more general background cosmological models \citep{grillo_cosmography_2024}.
%
SN H0pe is of particular interest as the first lensed SN\,Ia to provide an $H_0$ result competitive with local measurements, resulting in $75.4^{+8.1}_{-5.5}$ km s$^{-1}$ Mpc$^{-1}$ \citep{pascale_sn_2024}. 

SNe\,Ia are of particular value when strongly lensed, as their standardizable absolute brightness can provide additional leverage for lens modeling by limiting the uncertainty caused by the mass-sheet degeneracy \citep{falco_model_1985,kolatt_gravitational_1998,holz_seeing_2001,oguri_gravitational_2003,patel_three_2014,nordin_lensed_2014,rodney_illuminating_2015,xu_lens_2016,foxley-marrable_impact_2018,birrer_hubble_2022}, though only in cases where millilensing and microlensing are not extreme \citep[see][]{goobar_iptf16geu_2017,yahalomi_quadruply_2017,foxley-marrable_impact_2018,dhawan_magnification_2019}. Additionally, SNe\,Ia have well-understood models of light curve evolution \citep{hsiao_k_2007,guy_supernova_2010,saunders_snemo_2018,leget_sugar_2020,kenworthy_salt3_2021,mandel_hierarchical_2022,pierel_salt3-nir_2022} that enable precise time-delay measurements using color curves, which removes the effects of macro- and achromatic microlensing \citep[e.g.,][]{pierel_turning_2019,huber_holismokes_2021,rodney_gravitationally_2021,pierel_lenswatch_2023}. Both iPTF16geu \citep{goobar_iptf16geu_2017} and SN Zwicky \citep{goobar_uncovering_2023,pierel_lenswatch_2023} were SNe\,Ia, but each had very short time-delays of $\sim0.25$-$1.5$ days that precluded competitive $H_0$ measurements \citep{dhawan_magnification_2019,pierel_lenswatch_2023}.

``SN Requiem'' \citep{rodney_gravitationally_2021} was the first cluster-scale \textit{photometrically} classified multiply-imaged SN\,Ia, appearing at  $z=1.95$ in the MRG-M0138 galaxy that is lensed by the MACS J0138.0$-$2155 cluster (Figure \ref{fig:color_im}). Unfortunately the SN was found archivally several years after it had faded, precluding an $H_0$ measurement with the visible images. However, there is a fourth image of SN Requiem expected in $\sim2035$, an astounding $\sim20$ year time delay \citep{rodney_gravitationally_2021}. In a truly remarkable turn of events, a second lensed SN\,Ia (dubbed \lensedsn) has been found by the \textit{James Webb Space Telescope} (\textit{JWST}) in the same host galaxy as SN Requiem at $z=1.95$. The discovery visit was on 2023 November 17 (\textit{JWST}-GO-2345, PI Newman), which included \textit{JWST} NIRCam imaging and NIRSpec IFU observations. Previous observations of MACS J0138.0$-$2155 and MRG-M0138 are described in \citet[][hereafter N18]{newman_resolving_2018}, and SN Requiem is described in \citet{rodney_gravitationally_2021}. 

This work presents the discovery and observations of \lensedsn, and is the start of a series of papers analyzing the system. The outline of this paper is as follows: Section \ref{sec:obs} summarizes the \textit{Hubble Space Telescope} (\textit{HST}) and \textit{JWST} observations taken to follow \lensedsn. Section \ref{sec:encore} leverages the spectrum from Section \ref{sec:obs} to identify \lensedsn as an SN\,Ia, while Section \ref{sec:host} discusses our expectations surrounding the discovery of a second lensed SN\,Ia in the same host galaxy. We conclude with a discussion of the implications for \lensedsn in Section \ref{sec:conclusion}. 

\begin{figure}
    \centering
    \includegraphics[trim={0.5cm 0cm 2cm 0cm},clip,width=\columnwidth]{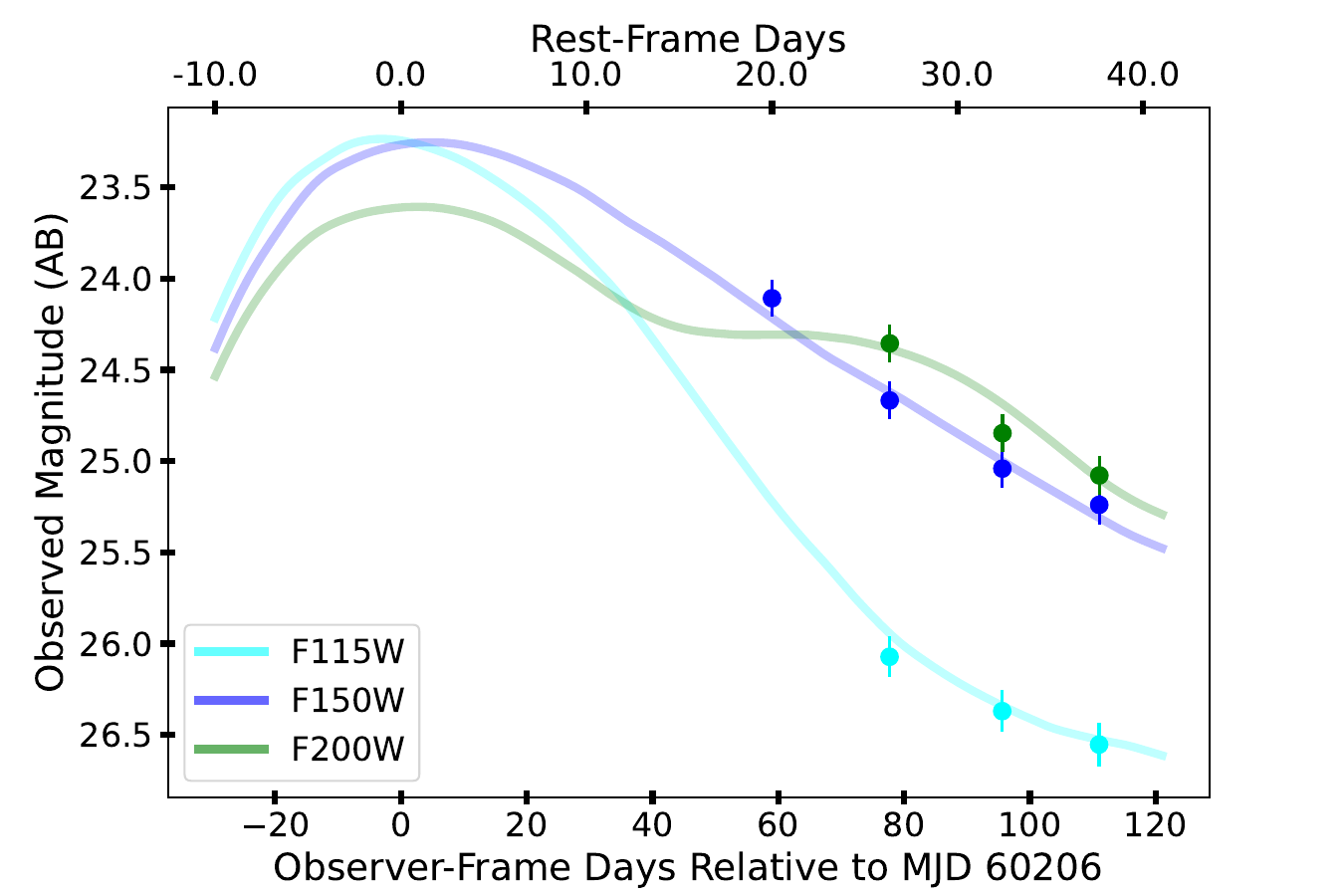}
    \caption{The observed light curve in the \textit{JWST}/NIRCam short-wavelength filters for image A, the brightest and last to arrive (see Figure \ref{fig:cutouts}). The long-wavelength filters (F277W, F356W, F444W) were observed but a more in-depth host galaxy modeling effort (or a template image) will be required to measure accurate photometry. }
    \label{fig:lc}
\end{figure}

\section{Summary of Observations}
\label{sec:obs}

\begin{figure*}[!t]
    \includegraphics[trim={6.5cm 10cm 5cm 10cm},clip, width=\linewidth]{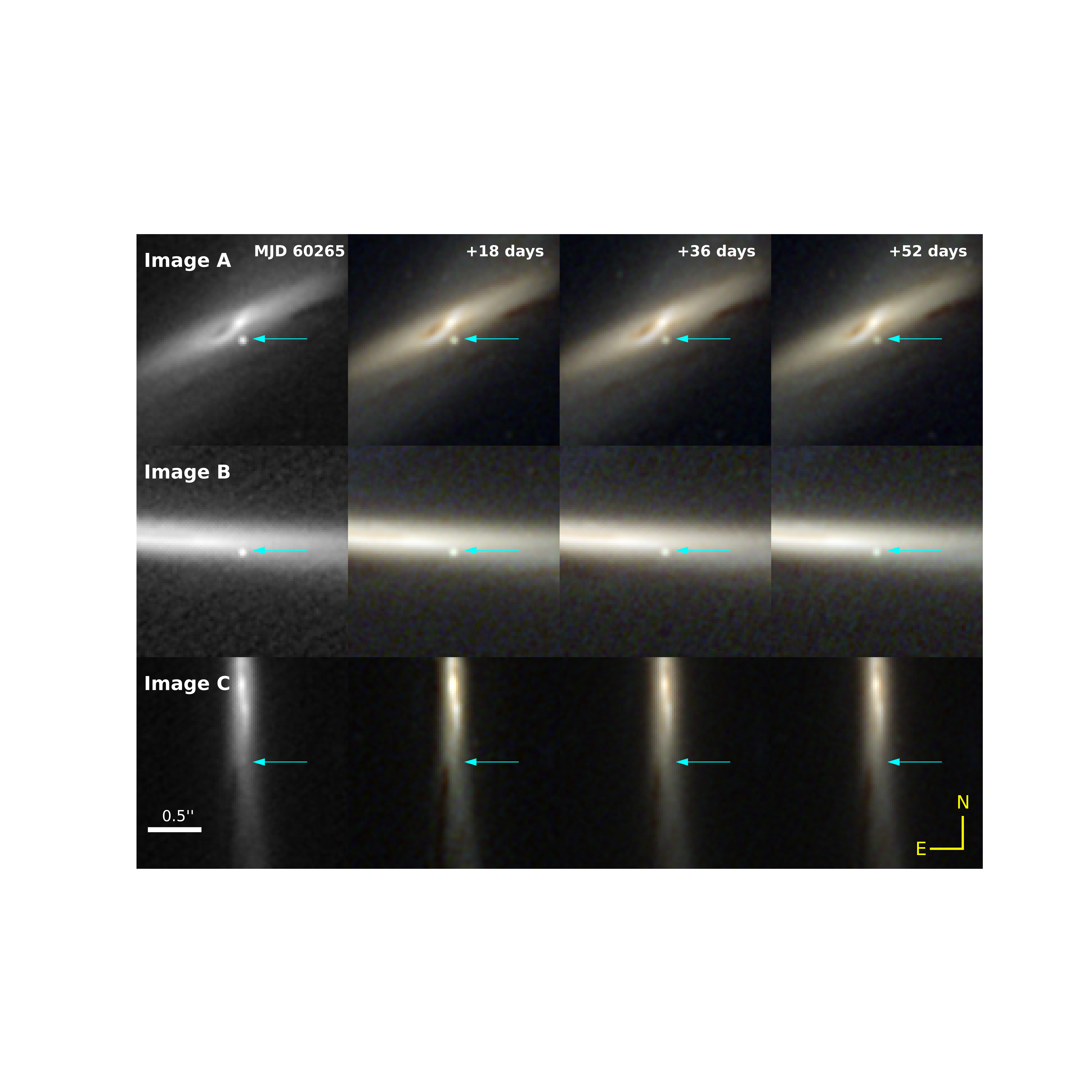}
    \caption{\label{fig:cutouts}\textit{JWST}/NIRCam color cutouts with F115W (blue), F150W (green) and F200W (red) centered on image A (top), B (middle), and C (bottom) of \lensedsn. Note the discovery program only detected \lensedsn in F150W, so the first column is a grey-scale image in F150W, while the remainder are 3-color images. The images were drizzled to $0\arcsec.02/$pix, with the image scale, orientation, and modified julian date (MJD; relative to discovery) are shown. \lensedsn is brightest in the discovery epoch and is on the decline in the subsequent epochs, but the evolution is relatively small at this redshift and is therefore difficult to see by eye (see Figure \ref{fig:lc}). Image C was first to arrive followed by image B, and although image C is not clearly visible by eye, it was detected at $\sim5\sigma$ (See Section \ref{sub:obs_jwst_img}). }
\end{figure*}

\lensedsn was discovered by the \textit{JWST}-GO-2345 team in F150W NIRCam imaging taken 2023 November 17 (MJD $60265$) by comparing the data with an archival \hst WFC3/IR F160W image (2016 July 18, MJD $57587$; N18). These two filters are extremely well-matched in wavelength and transmission, and the source was very bright ($\lesssim24$\,mag~AB in both images A and B, see Figure \ref{fig:cutouts}) in F150W but absent in F160W. However, the separation of \lensedsn from its host galaxy MRG-M0138 is only $\sim0\arcsec.1$, which is roughly the pixel-scale of the \textit{HST} WFC3/IR imaging. We therefore also compared the point-source positions and relative brightnesses to lens model predictions (Ertl et al., Suyu et al. in preparation), and confirmed with forced photometry that there was an increase in flux corresponding to the apparent brightness of \lensedsn between the F160W and F150W imaging. All tests were in agreement that this was indeed a multiply-imaged SN, with (at least) two images of the SN detected (images A and B, see Figure \ref{fig:cutouts}). Further efforts to remove host galaxy light near image C revealed that a third image of \lensedsn was also detected. There two additional images of the host galaxy located in the northeast of the cluster (Figure \ref{fig:color_im}), with the next image of \lensedsn expected from our lens modeling to arrive in about a decade (Ertl et al.~in preparation). The sharp radial profile of images D and E for MRG-M$0138$ will provide relatively tight constraints on these long time delays, which will enable a targeted follow-up campaign similar to SN Refsdal \citep{kelly_deja_2016}. N18 reported a spectroscopic redshift for MRG-M0138 of $z=1.95$, and the host properties alone suggested that \lensedsn was likely an SN\,Ia at $z=1.95$ \citep{rodney_gravitationally_2021}. 

\begin{table*}[!t]
    \centering
    \caption{\label{tab:obs} Summary of observations taken of SN Encore.}
    
    \begin{tabular*}{\linewidth}{@{\extracolsep{\stretch{1}}}*{7}{c}}
\toprule
Program ID&Obs. Type&Telescope&Instrument&MJD&\multicolumn{1}{c}{Filter/Grating}&\multicolumn{1}{c}{Exp. Time (s)}\\
\hline
2345&Imaging&\textit{JWST}&NIRCam&$60265$&F150W&$773$\\
2345&Imaging&\textit{JWST}&NIRCam&$60265$&F444W&$773$\\
2345&Spectroscopy&\textit{JWST}&NIRSpec IFU&$60266$&G235M&$7586$\\
2345&Spectroscopy&\textit{JWST}&NIRSpec IFU&$60305$&G140M&$8170$\\

\hline
6549&Imaging&\textit{JWST}&NIRCam&$60283$&F115W&$1417$\\
6549&Imaging&\textit{JWST}&NIRCam&$60283$&F150W&$859$\\
6549&Imaging&\textit{JWST}&NIRCam&$60283$&F200W&$859$\\
6549&Imaging&\textit{JWST}&NIRCam&$60283$&F277W&$859$\\
6549&Imaging&\textit{JWST}&NIRCam&$60283$&F356W&$859$\\
6549&Imaging&\textit{JWST}&NIRCam&$60283$&F444W&$1417$\\
6549&Imaging&\textit{JWST}&NIRCam&$60301$&F115W&$1589$\\
6549&Imaging&\textit{JWST}&NIRCam&$60301$&F150W&$1074$\\
6549&Imaging&\textit{JWST}&NIRCam&$60301$&F200W&$1074$\\
6549&Imaging&\textit{JWST}&NIRCam&$60301$&F277W&$1074$\\
6549&Imaging&\textit{JWST}&NIRCam&$60301$&F356W&$1074$\\
6549&Imaging&\textit{JWST}&NIRCam&$60301$&F444W&$1589$\\
6549&Imaging&\textit{JWST}&NIRCam&$60317$&F115W&$1589$\\
6549&Imaging&\textit{JWST}&NIRCam&$60317$&F150W&$1074$\\
6549&Imaging&\textit{JWST}&NIRCam&$60317$&F200W&$1074$\\
6549&Imaging&\textit{JWST}&NIRCam&$60317$&F277W&$1074$\\
6549&Imaging&\textit{JWST}&NIRCam&$60317$&F356W&$1074$\\
6549&Imaging&\textit{JWST}&NIRCam&$60317$&F444W&$1589$\\
\hline
16264&Imaging&\textit{HST}&WFC3/IR&$60294$&F105W&$6913$\\
16264&Imaging&\textit{HST}&WFC3/IR&$60294$&F125W&$4609$\\
16264&Imaging&\textit{HST}&WFC3/IR&$60294$&F160W&$2304$\\
16264&Imaging&\textit{HST}&WFC3/IR&$60339$&F105W&$9318$\\
16264&Imaging&\textit{HST}&WFC3/IR&$60339$&F125W&$4609$\\
16264&Imaging&\textit{HST}&WFC3/IR&$60339$&F160W&$2304$\\
\hline

\hline
    \end{tabular*}
\begin{flushleft}
\tablecomments{Columns are: \textit{JWST}/\textit{HST} Program ID, observation type, telescope name, instrument name, Modified Julian date, filter/grating, and exposure time in seconds.}

\end{flushleft}
\end{table*}

\subsection{\textit{JWST} Data}
\label{sub:obs_jwst}
\subsubsection{Imaging}
\label{sub:obs_jwst_img}
An \textit{HST} program was triggered soon after the discovery of \lensedsn, which is described in the following section. However, \textit{JWST} was still required for two reasons: 1) the small host separation described above meant that even with difference imaging, it would be very difficult to detect \lensedsn in any but the brightest filter (F160W for \textit{HST}), and 2) based on the discovery epoch \lensedsn was between peak brightness and the second infrared (IR) maximum for SNe\,Ia \citep{hsiao_k_2007,krisciunas_carnegie_2017,pierel_salt3-nir_2022}. At $z=1.95$ \textit{HST} is only able to cover rest-frame uBV filters, meaning \textit{JWST} was needed to reach the rest-frame near-IR and leverage the second infrared maximum for time-delay measurements.

A disruptive \webb director's discretionary time (DDT) proposal was subsequently approved (\textit{JWST}-GO-6549, PI Pierel), which provided three additional epochs of NIRCam imaging in six filters (Table \ref{tab:obs}). The first NIRCam epoch occurred on 2023 December 5 (MJD $60283$), $18$ ($\sim6$ rest-frame) days after discovery. The next two epochs took place with a cadence of $\sim17$ observer-frame ($\sim6$ rest-frame) days, on 2023 December 23 (MJD $60301$) and 2024 January 8 (MJD $60317$). This meant the \textit{JWST} light curve for each image contains four observations, each separated by $\sim6$ rest-frame days. Although the image C of \lensedsn is too faint to see clearly by eye in Figure \ref{fig:cutouts}, it is $\sim27$ AB mag in F150W and detected at $\sim5\sigma$ with careful subtraction of the host galaxy light profile in \textit{JWST} imaging. The full \textit{JWST} dataset therefore effectively produces a SN\,Ia light curve with $12$ epochs and rest-frame UV-NIR wavelength coverage, which is sufficient to accurately measure phases of each image and therefore the relative time delays \citep[e.g.,][]{pierel_turning_2019}. The  summary of observations is given in Table \ref{tab:obs}.  

All the NIRCam data were retrieved from the STScI MAST\footnote{\url{https://mast.stsci.edu}} archive, and were processed and calibrated using the JWST Pipeline\footnote{\url{https://github.com/spacetelescope/jwst}} 
\citep{bushouse_jwst_2022} version 1.12.5, with CRDS reference files defined in 1180.pmap, with additional improvements described here as follows. As part of the initial processing with the Stage 1 portion of the pipeline, additional corrections were applied to remove 1/f noise as well as low-level background variations across the detectors, including the removal of wisps and other low-level artifacts, with these techniques all described in more detail in \citet{windhorst_jwst_2023}. The Stage 2 portion of the pipeline was then run to apply photometric calibration to physical units, using the latest photometric calibrations (Boyer et al. 2023).
%
%
%
The images were then all astrometrically aligned directly to the Gaia-DR3\footnote{\url{https://www.cosmos.esa.int/web/gaia/dr3}} reference frame and finally combined into mosaics for each filter with all distortion removed, using the Stage 3 portion of the pipeline, where these mosaics were constructed for each of the individual epochs, as well as full-depth versions combining all the epochs.

\subsubsection{Spectroscopy}
\label{sub:obs_jwst_spec}
The \textit{JWST}-GO-2345 program also included NIRSpec integral field unit (IFU) spectroscopy for image A of MRG-M0138 using both the G140M and G235M gratings. There was a failure in the G140M observation causing it to be scheduled $39$ days later on 2023 December 27, but the G235M observation was successful. Here we extract the SN spectrum for the G235M grating, aligned to the G140M spectrum using their overlapping wavelength range, with the G140M spectrum anchored to the F150W NIRCam image to reveal the SN location. The SN spaxels have unique spectral features in the G235M observation, distinct from the galaxy spectra, making it possible to disentangle. We establish contour levels for each IFU datacube by calculating the median flux across the entire wavelength range, which allows us to subtract the host galaxy spectra from the spaxels containing SN flux.  Finally, we isolate the SN spectrum by subtracting off the median spectrum of host galaxy from what is observed in the SN spaxels. We leave discussion of the G140M spectrum and in-depth analysis of the G235M spectrum to the upcoming SN spectroscopic analysis (Dhawan et al. in preparation), host galaxy analysis (Newman et al. in preparation), and investigations into relationships between the host galaxy and SN\,Ia properties (Larison et al. in preparation). The G235M spectrum is used briefly here to simply confirm the spectroscopic SN type for \lensedsn in Section \ref{sec:encore}, and is shown in Figure \ref{fig:spec}.

\subsection{\textit{HST} Data}
\label{sub:obs_hst}
LensWatch\footnote{\url{https://www.lenswatch.org}} is a collaboration with the goal of finding
gravitationally lensed SNe, both by monitoring active transient surveys \citep[e.g.,][]{fremling_zwicky_2020,jones_young_2021} and by
way of targeted surveys \citep{craig_targeted_2021}. The collaboration
maintained a Cycle 28 \textit{HST} program (HST-GO-16264) given long-term (three-cycle) target-of-opportunity (ToO) status. The program included three ToO
triggers (two non-disruptive, one disruptive), and was designed
to provide high-resolution follow-up imaging for a ground-based lensed SN discovery, which is critical for galaxy-scale
multiply imaged SNe due to their small image separations \citep[e.g.,][]{goobar_iptf16geu_2017}. LensWatch triggered \textit{HST}-GO program 16264 on 2023 November 20 (MJD $60268$, three days after discovery) to obtain follow-up imaging of \lensedsn, including $14$ orbits in three filters spread over two epochs. The trigger was disruptive but there were small delays related to \textit{HST} being in safe mode, which led to the first epoch being $26$ ($\sim9$ rest-frame) days later on 2023 December 16 (MJD $60294$) and the second epoch $45$ days after the first on 2024 January 30 (MJD $60339$). These observations are separated by $10$-$22$ observer-frame ($3$-$7$ rest-frame) days from the \textit{JWST} observations described in the previous section, and were designed to be complementary to \textit{JWST}. All observations are summarized in Table \ref{tab:obs}.

All the HST data were retrieved from the STScI MAST archive, and the calibrated exposures were subsequently processed with additional improvements described here as follows. The default astrometry from the archive was corrected by realigning all the HST exposures to one another, as well as to the JWST images and to Gaia-DR3, using an updated version of the HST ``mosaicdrizzle'' pipeline first described in \citet{koekemoer_candels_2011}, which also removes residual low-level background variations across the detectors. This pipeline was then used to combine all the HST data into mosaics for each filter with all distortion removed, producing mosaics for the separate epochs as well as full-depth versions combining all the epochs.

\section{Classifying SN Encore}
\label{sec:encore}
\begin{figure}[!h]
    \centering
    \includegraphics[trim={.5cm 0cm 1.5cm 1cm},clip,width=\linewidth]{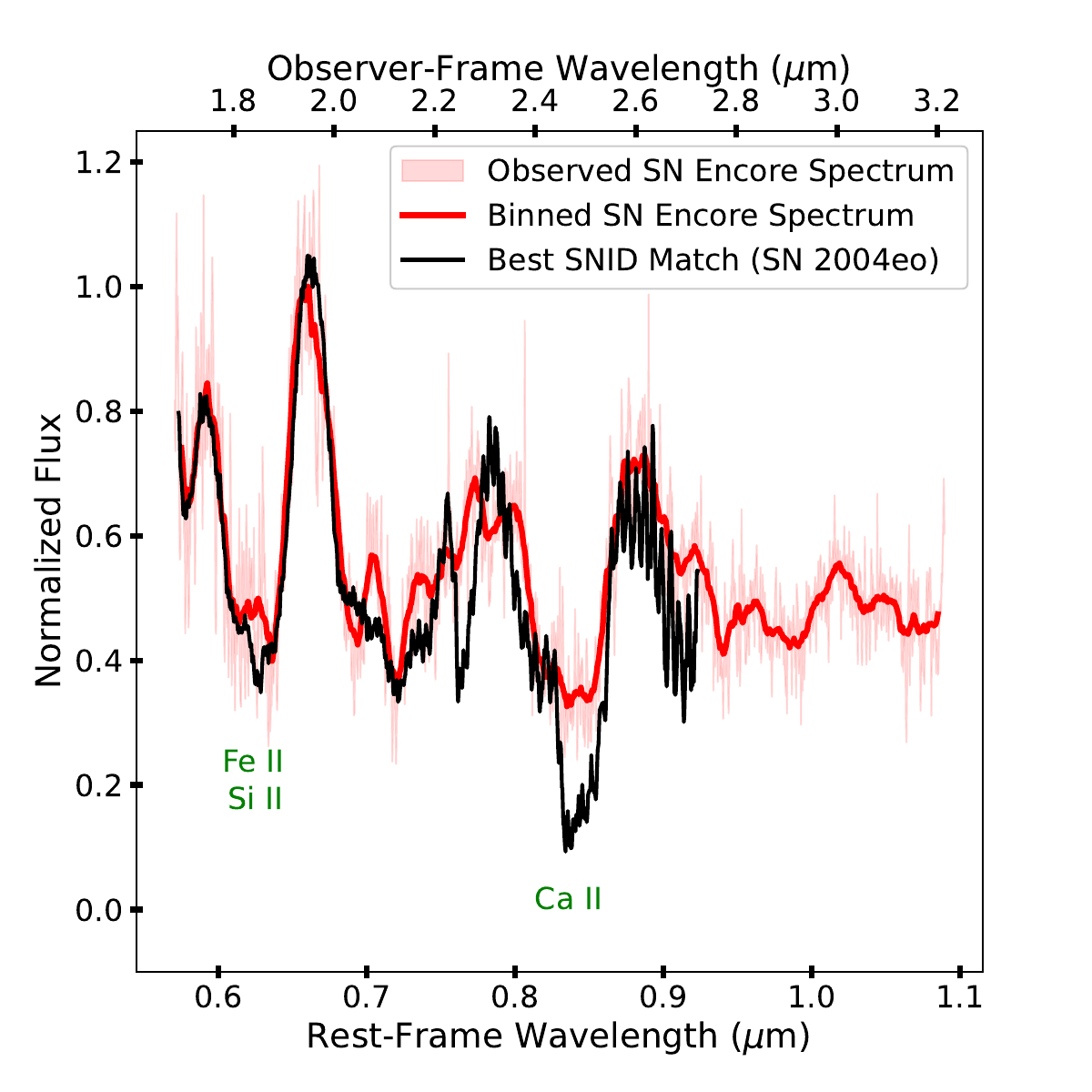}
    \caption{The (normalized) extracted G235M spectrum of \lensedsn with $1\sigma$ uncertainty (faint red), with the binned spectrum superimposed (thick red line). The best-match spectrum from SNID (SN 2004eo; black line) is shown for comparison, with characteristic SN\,Ia features labeled in green text, confirming that \lensedsn is of Type Ia.}
    \label{fig:spec}
\end{figure}

Lensed SNe\,Ia offer smaller uncertainties for $H_0$ than lensed core-collapse SNe (CC\,SNe) \citep[e.g.,][]{goldstein_precise_2018,pierel_turning_2019,birrer_hubble_2022} and unique leverage on lens modeling systematics \citep{pascale_sn_2024}, making them the most valuable type of lensed SN. We infer the type of \lensedsn using the supernova identification \citep[SNID;][]{blondin_determining_2007} and Next-Generation SuperFit (NGSF)\footnote{\href{https://github.com/oyaron/NGSF}{https://github.com/oyaron/NGSF}} packages with the observed G235M spectrum (Table \ref{tab:obs}, Figure \ref{fig:spec}). Both packages classify \lensedsn as Type Ia with $>90\%$ confidence ($9$ of the top $10$ best-fit reference spectra are from SNe\,Ia and the last is a SN\,Ia-91T, which is a subclass known to be spectroscopically similar to normal SNe\,Ia at late times), with the best matches being SN~2004eo (SNID) and SN~2007sr (NGSF). Both of these SNe decline faster than the mean decline rate (SALT3 $x_1\lesssim0$), which is in agreement with light curve fitting for \lensedsn ($x_1\sim-1.3$), but all three SNe exhibit light curve shapes that are well-within traditional cosmology cuts \citep[e.g., $-3<x_1<3$][]{scolnic_complete_2018}.  NGSF also returns an estimate of the host galaxy contamination in the SN spectrum, which is $<1\%$. Finally, the phase is constrained by the classification codes to $26.5\pm5.7$ rest-frame days, which is in $1.1\sigma$ agreement with the results of our light curve fitting ($20.1\pm3$ rest-frame days; Figure \ref{fig:lc}). Overall the result of this analysis is therefore that \lensedsn is a SN\,Ia, and moreover is likely a relatively fast-declining SN that should nevertheless be standardizable. A full analysis of both the G235M and G140M spectrum, including a comparison to low-$z$ SN\,Ia spectra and theoretical models for SN\,Ia progenitors, will be presented in Dhawan et al. (in preparation). 

While the most precise time-delay measurements come from lensed SNe observed before peak brightness, a lensed SN\,Ia observed roughly in the phase of images A and B in multiple filters should still provide a time-delay measurement uncertainty of $\sim5$ observer-frame days \citep{pierel_turning_2019}. Image C was observed well after peak brightness in the nebular phase of SNe\,Ia, where the slow and uncertain evolution will likely result in a lower time-delay precision of $\gtrsim10$ observer frame days \citep{pierel_jwst_2024}. Initial expectations from lens models of the MACS J0138.0$-$2155 cluster and \lensedsn photometry suggest the delay between images A and B will be $\gtrsim1$ month, while the delay between images A and C will be $\sim1$ year. Assuming a fiducial cluster lens model uncertainty of $\sim6\%$ \citep{grillo_accuracy_2020,kelly_constraints_2023}, \lensedsn should produce an $H_0$ uncertainty of $\sim10\%$. A detailed lens modeling analysis will be presented by Ertl et al. and Suyu et al. (in preparation), while measurements of the time-delay and $H_0$ will be presented by Pierel et al. (in preparation).

\section{Two Lensed SNe\,Ia from a Single Host Galaxy}
\label{sec:host}
While it seems remarkable to find two lensed SNe\,Ia in the same host galaxy at $z=1.95$, we briefly explore the probability of such a discovery. A full analysis of the relationship between the local/global properties of MRG-M0138 and both \lensedsn and Requiem will be presented by Larison et al. (in preparation), leveraging the full dataset shown in Table \ref{tab:obs}. Here, we simply provide estimates assuming the results of N18.

N18 describes the specifics of their data acquisition, and here we use their host properties derived from a near-IR spectrum from the MOSFIRE spectrograph instrument on the Keck I telescope. N18 have already provided best-fit SFHs by fitting the MRG-M0138 spectrum with a parametric SFH model. The parametric model they use is of the form, $\mathrm{SFR \propto exp}(-t/\tau$). Table 5 in N18 provides the $\tau$ value for the SFH based on two images of MRG-M0138 -- we adopt an average value of $180$ Myr which is consistent with fits from both images.

There are one of two approaches we investigate here. In the first approach, we assume the SN\,Ia delay-time distribution (DTD) to be an exponential function from \citet{strolger_delay_2020}, and convolve it with our SFHs derived above to obtain an SN\,Ia rate for MRG-M0138. The SN\,Ia rate calculation also requires assumptions to be made about the form of the initial mass function (IMF) and the fraction of stars within 3--8 M$_\odot$ that are ultimately successful SN\,Ia explosions ($\epsilon$). Following \citet{strolger_delay_2020}, we assume an IMF with a power-law index of $\alpha\,\approx\,-2.3$ \citep{salpeter_luminosity_1955,kroupa_variation_2001} and $\epsilon{=}0.06$. 

MRG-M0138 is quiescent and not exceptionally star forming in the observed epoch, at a rate between $\approx 1.8$ and 2.6 ($\pm1.7$) $M_{\odot}$ yr$^{-1}$, depending on which image is used. Using the method above implies a SN\,Ia rate of approximately 90 events (1000 yr)$^{-1}$, which is consistent with the average rate of SNe~Ia per galaxy, regardless of mass, from \citet{li_nearby_2011}. Assuming effectively a 10-year monitoring of the host, i.e., in which no detectable SN\,Ia event was missed in the last 10 years (3.39 years in the rest frame of the host), it would be reasonable to expect the probability of having one SN\,Ia in that time frame to be low, but not zero, with $\rm{Pois}(k=1)=0.23$. Similarly, the probability of finding two such events is also low, with $\rm{Pois}(k=2)=0.03$. It is interesting to note that assuming a power-law delay time distribution \citep[e.g.,][]{maoz_delay-time_2012, freundlich_delay_2021} would make the rate at the observed epoch even lower, 5.3 events (1000 yr)$^{-1}$, and the likelihood of observing two supernovae in MRG-M0138 in 10 years, $\rm{Pois}(k=2)=0.0001$.

However, another approach could be to estimate the expected yield from the mass-weighted rates as is done in Smith et al. (2020) and Scolnic et al. (2020). This has the convenience of estimating the rate from the galaxy's current properties, but loses the detailed analysis provided by the Strolger et al. (2020) approach. Li et al. (2011) show an average mass-weighted SN Ia rate per galaxy of $0.54\pm0.12$ events per century per $10^{10}$ M$_{\odot}$. 
N18 determine $\log{\rm M}_{*}/{\rm M}_{\odot}=11.7\pm0.2$ (for both images of the host), resulting in a expected rest-frame yield of about 260 (1000 yr)$^{-1}$, or 0.88 events in the 10-year observation window. The probability of having one SN\,Ia in that time-frame would therefore be a bit higher, with $\rm{Pois}(k=1)=0.37$, and the probability of finding two such events seemingly more feasible, at $\approx16\%$. Both approaches are broadly consistent with the rate of discovery of SN~Ia siblings from \citet{kelsey_archival_2024} and \citet{scolnic_supernova_2020}.


\section{Discussion}
\label{sec:conclusion}
Here we presented the discovery and observations of \lensedsn, a SN lensed by the MACS J0138.0$-$2155 cluster and hosted by the galaxy MRG-M0138 at $z=1.95$. We spectroscopically confirm that \lensedsn is of Type Ia, making it just the second lensed SN\,Ia and third overall lensed SN useful for measuring $H_0$ with $\lesssim10\%$ uncertainty. Remarkably, \lensedsn is the second lensed SN\,Ia hosted by MRG-M0138 in the last $\sim7$ years, making it the first known case of two lensed SNe from the same host galaxy. While the appearance of two SN Ia in the same multiply imaged galaxy is remarkable, MRG-M0138 was targeted because it is by far the brightest and most massive ($M_*= 5\times10^{11} M_\odot$) among the rare class of magnified quiescent galaxies at $z \gtrsim 2$ \citep{toft_massive_2017,akhshik_requiem-2d_2023}. MRG-M0138's extreme stellar mass, old stellar population, and lack of strong dust attenuation lead to a low but non-negligible derived probability of $\lesssim3\%$ for detecting two SNe\,Ia in the last $\sim$decade. SN Requiem and \lensedsn will both reappear in the next $\sim$$5$-$10$ years, providing an exceedingly rare opportunity to monitor MACS J0138.0$-$2156 for two SNe\,Ia with known reappearance location and time. The resulting baseline will be unprecedented for time-delay cosmography and can provide unique insights into SN\,Ia physics at $z=1.95$.

While \lensedsn will not break the $H_0$ tension \citep[e.g.,][]{riess_comprehensive_2022} on its own, it still plays a critical role as the third lensed SN (and second lensed SN\,Ia) to provide a precise measurement of $H_0$. With a sample of three the combined $H_0$ uncertainty could plausibly be $\sim5\%$, and we will start to have a clearer picture of the relative agreement across lensed SN measurements of $H_0$. Upcoming surveys such as the Vera C. Rubin Observatory Legacy Survey of Space and Time \citep[LSST;][]{ivezic_lsst_2019,arendse_detecting_2023} and the \textit{Nancy Grace Roman Space Telescope} High Latitude Time Domain Survey \citep[HLTDS;][]{pierel_projected_2021,rose_reference_2021} are expected to deliver dozens of useful galaxy-and cluster-scale multiply-imaged SNe, and although more complicated, clusters provide longer time-delays, wider image separations, and extra lens modeling constraints on the mass-sheet degeneracy \citep{falco_model_1985,grillo_accuracy_2020, grillo_cosmography_2024, pascale_sn_2024}.  As a result, cluster-lensed SNe are likely to continue as an extremely valuable subset of the future cosmological sample.




\clearpage

\begin{center}
    \textbf{Acknowledgements}
\end{center}

This paper is based in part on observations with the NASA/ESA Hubble Space Telescope and James Webb Space Telescope obtained from the Mikulski Archive for Space Telescopes at STScI. We thank the DDT and JWST/HST scheduling teams at STScI for extraordinary effort in getting the DDT observations used here scheduled quickly. The specific observations analyzed can be accessed via \dataset[DOI: 10.17909/snj9-an10]{https://doi.org/10.17909/snj9-an10}"; support was provided to JDRP and ME through program HST-GO-16264. JDRP is supported by NASA through a Einstein
Fellowship grant No. HF2-51541.001, and AJS  by HF2-51492 awarded by the Space
Telescope Science Institute (STScI), which is operated by the
Association of Universities for Research in Astronomy, Inc.,
for NASA, under contract NAS5-26555.
RAW acknowledges support from NASA \textit{JWST} Interdisciplinary Scientist grants NAG5-12460, NNX14AN10G and 80NSSC18K0200 from GSFC. 
SHS thanks the Max Planck Society for support through the Max Planck Fellowship. This project has received funding from the European Research Council (ERC) under the European Union’s Horizon 2020 research and innovation programme (LENSNOVA: grant agreement No 771776). This research is supported in part by the Excellence Cluster ORIGINS which is funded by the Deutsche Forschungsgemeinschaft (DFG, German Research Foundation) under Germany's Excellence Strategy -- EXC-2094 -- 390783311. X.H. acknowledges the University of San Francisco Faculty Development Fund. SD acknowledges support from the Marie Curie Individual Fellowship under grant ID 890695 and a Junior Research Fellowship at Lucy Cavendish College. C.G. and G.G. acknowledge support through grant PRIN-MIUR 2020SKSTHZ. FP acknowledges support from the Spanish Ministerio de Ciencia, Innovación y Universidades (MICINN) under grant numbers PID2019-110614GB-C21 and PID2022-141915NB-C21. Support for program GO-2345 was provided by NASA through a grant from the Space Telescope Science Institute, which is operated by the Association of Universities for Research in Astronomy, Inc., under NASA contract NAS 5-03127. K.G. acknowledges support from Australian Research Council Laureate Fellowship FL180100060.
This work was supported by research grants (VIL16599, VIL54489) from VILLUM FONDEN.
RAW acknowledges support from NASA JWST Interdisciplinary Scientist grants
NAG5-12460, NNX14AN10G and 80NSSC18K0200 from GSFC. Support for program JWST GO-04446 was provided by NASA through a grant from the Space Telescope Science Institute, which is operated by the Association of Universities for Research in Astronomy, Inc., under NASA contract NAS 5-26555. SB is supported by ERC StG 101076080.SS has received funding from the European Union’s Horizon 2022 research and innovation programme under the Marie Skłodowska-Curie grant agreement No 101105167 — FASTIDIoUS. CG is supported by a VILLUM FONDEN Young Investigator Grant (project number 25501). P. L. K. acknowledges support from NSF AAG-2308051. C.L. acknowledges support from the National Science Foundation Graduate Research Fellowship under grant No. DGE-2233066. RAW, SHC, and RAJ acknowledge support from NASA JWST Interdisciplinary Scientist grants NAG5-12460, NNX14AN10G, and 80NSSC18K0200 from GSFC. This work has received funding from the European Research Council (ERC) under the European Union’s Horizon 2020 research and innovation programme (LensEra: grant agreement No. 945536). TC is funded by the Royal Society through a University Research Fellowship. Part of this research was carried out at the Jet Propulsion Laboratory, California Institute of Technology, under a contract with the National Aeronautics and Space Administration (80NM0018D0004). AZ acknowledges support by Grant No. 2020750 from the United States-Israel Binational Science Foundation (BSF) and Grant No. 2109066 from the United States National Science Foundation (NSF); by the Ministry of Science \& Technology, Israel; and by the Israel Science Foundation Grant No. 864/23.

\clearpage

\bibliographystyle{aasjournal}


\end{document}